\documentclass[11pt,onecolumn,amssymb,nofootinbib]{revtex4}
\usepackage{amsmath, amsthm, amscd, amssymb}
\usepackage{graphicx, braket}
\usepackage{bm}
\usepackage{bbm}
\usepackage{epstopdf}

\begin{document}

\title{\bf A mechanism for a ``leaky'' black hole to catalyze galaxy formation} \bigskip
\leftline{Essay written for the Gravity Research Foundation 2022 Awards for Essays on Gravitation;}
\leftline{submitted March 18, 2022.}

\author{Stephen L. Adler}
\email{adler@ias.edu} \affiliation{Institute for Advanced Study,
Einstein Drive, Princeton, NJ 08540, USA.}

\begin{abstract}
In the gravitational field of a Schwarzschild-like black hole,  particles infalling from rest at infinity, and black hole ``wind'' particles with relativistic velocity leaking radially out from the nominal horizon, both have the same magnitude of velocity at any radius from the hole.  Hence when equally massive infalling and wind particles collide at any radius, they yield collision products  with zero center of mass radial velocity, which can then nucleate star formation at the collision radius.    We suggest that this gives a mechanism by which a central black hole can catalyze galaxy formation.  For disk galaxies,  this mechanism explains the observed  approximately exponential falloff of the surface brightness with radius, and gives an estimate of the associated scale length.

\end{abstract}

\maketitle

A number of salient facts suggest that galaxies and black holes are intimately related. It is now suspected that every galaxy harbors a black hole at its center, and it is known that the central bulge size of galaxies is correlated with the mass of the internal black hole. These facts,    as well as observational data \cite{V1},\cite{V2},\cite{V3},  suggest  that the black hole may be formed first and then plays a role in galaxy formation, but a concrete mechanism has been lacking.

From a different direction,  as reviewed in \cite{A1}, we have studied consequences of the postulate that ``dark energy'' arises from an action constructed from
nonderivative metric components so as to be invariant under the Weyl scaling $g_{\mu\nu}(x) \to \lambda(x) g_{\mu\nu}(x)$, where $\lambda(x)$ is a general scalar function.  This novel dark energy action, which reduces to the usually assumed one in cosmological contexts when $g_{00}(x)=1$, is given by
\begin{equation}\label{dark}
S_{\rm dark~energy}=-\frac{\Lambda}{8 \pi G} \int d^4x  ({}^{(4)}g)^{1/2}(g_{00})^{-2}~~~,
\end{equation}
where $\Lambda$ is the cosmological constant, $G$ is Newton's constant, and ${}^{(4)}g=-\det({g_{\mu\nu})}$.
Black holes as modified by this action
differ from their standard general relativity form within a distance of order $2 \times 10^{-18} (M/M_\odot)^2$cm from a nominal horizon radius $2M$, with $M$ the black hole mass and $M_\odot$ the solar mass.    Moreover,  detailed calculations in the spherically symmetric case show that as a consequence of the $(g_{00})^{-2}$ factor in Eq. \eqref{dark}, which would become infinite if $g_{00}$ were to vanish,  these modified, ``Schwarzschild-like'' black holes have neither an event horizon \cite{F}  nor an apparent horizon \cite{A2}.\footnote{Schwarzschild-like black holes are a variant of the ``exotic compact objects'' reviewed in \cite{eco}, where the focus is on possible gravitational wave signatures stemming from absence of a horizon.} Thus, since there are no trapped surfaces,  they can be ``leaky'', which would allow a black hole wind that can influence processes outside the hole.  In a paper in preparation \cite{K}, we have studied a mechanism by which a black hole wind from Sgr A* can lead to the mysterious young star formation \cite{lu} near the central black hole of our galaxy.  In this essay, we suggest a mechanism by which a black hole wind can catalyze galaxy formation.

Our starting point is the general relativistic equation for an ingoing (or outgoing) radial timelike geodesic in the geometry of a Schwarzschild black hole, which using geometrized units with Newton's constant and the velocity of light both equal to unity takes the form,
\begin{equation}\label{geodesic}
 \left(\frac{dr}{d\tau}\right)^2-\frac{2M}{r}= \epsilon^2-1~~~.
\end{equation}
Here  $M$ is the hole mass and $\epsilon$ is the energy per unit mass of the ingoing or outgoing particle, and  we have multiplied the standard energy
formula by an overall  factor of 2.\footnote{As noted in \cite{G}, Eq. \eqref{geodesic}  is the same as the equation governing radial free fall in Newtonian gravity.}  Consider first
an infalling particle starting from rest at $r=\infty$, so that initially both terms on the left hand side of Eq. \eqref{geodesic} vanish, which implies that $\epsilon=1$.  At an arbitrary radius $R$, the infalling particle will then have velocity $v_{\rm in}$ given by
\begin{equation}\label{infall}
v_{\rm in}^2= \frac{2M}{R}~~~.
\end{equation}
When the particle reaches the nominal horizon of a Schwarzschild-like black hole, at $R=2M$, it will have velocity close to that of light.  Thus we expect the interior of the modified black hole to be populated by highly relativistic particles, making it plausible that those that leak out, which give rise to the black hole wind,  will have velocities close to that of light.

Consider next a black hole wind particle emerging radially from just outside the  Schwarzschild-like black hole nominal horizon at $r=2M$, with relativistic velocity $(dr/d\tau)^2 \simeq 1$.  Then both terms on the left hand side of Eq. \eqref{geodesic} are unity, which again implies that $\epsilon=1$.  At an arbitrary radius $R$, the outgoing wind particle will then have velocity $v_{\rm out}$ given by
\begin{equation}\label{outgo}
v_{\rm out}^2= \frac{2M}{R}~~~.
\end{equation}

Assuming now that the infalling and outgoing particles have the same mass, a collision of an infalling particle with an outgoing wind particle
will have zero center of mass radial velocity, and so will result in collision products that initially remain in the vicinity of radius $R$, and thus can nucleate star formation there. Components of velocity perpendicular to the radial direction can give the forming star angular momentum around the central black hole, so that under the attraction of the black hole it will migrate to a radius where it orbits the hole.    Since $R$ is arbitrary, this gives a mechanism by which stars can form over a wide range of radii from the central black hole, suggesting the following scenario for galaxy formation: First, gravitational collapse of dust particles leads to the formation of a black hole, which keeps growing as more dust (or later on a stray star) is accreted.  Collisions of wind particles from the hole
with infalling dust particles then nucleate star formation, so that as the hole grows in size, a galaxy of stars grows along with the hole.\footnote{We have
focused on the spherically symmetric case. When the hole is rotating the ratio of angular momentum to mass of the hole would be expected to
influence the geometry of the galaxy that is created.}

A simple collision length estimate for wind particles incident on intergalactic atomic hydrogen  is compatible with this mechanism.    Taking $\rho_H \simeq 1 \,{\rm m}^{-3}=10^{-6} {\rm cm}^{-3}$ as the density of hydrogen atoms in the proto-galactic region from which the galaxy forms, and
$\sigma_H \simeq \pi a_0^2 \simeq 10^{-16}{\rm cm}^2$, with $a_0$ the Bohr radius,  as the geometric scattering cross section for wind particles on hydrogen, we get a collision length $L_c=1/(\rho_H \sigma_H) \simeq 10^{22} {\rm cm}\simeq 3.2 {\rm kpc}$. Using the average distance between galaxies, one million light years $\simeq10^{24}{\rm cm}\simeq 3.1 \times 10^2 {\rm kpc}$,  as an estimate of the diameter of a proto-galactic region, there are $\sim 10^2$  collision lengths for a wind particle to cross the proto-galactic region, making a collision within this region very likely.

A more precise quantitative statement can be made. Our proposed mechanism for nucleation of star formation implies that the galactic star density profile as a function of the radius $R$ from the central hole will scale as a constant times
\begin{equation}\label{sph1}
 \exp(-R/L_c)~~~,
\end{equation}
since this gives the density profile of unscattered outgoing wind particles.
The effect of dissipation, combined with conservation of angular momentum from galactic rotation, leads to relaxation of an initially spherical galaxy to a disk.  Writing $R=(L^2+z^2)^{1/2}$ in terms of axial coordinates $L,z$, and integrating over $z$, the disk density profile for a thin disk corresponding to the spherical profile of Eq. \eqref{sph1} is
\begin{align}\label{sph2}
&\int_0^\infty dz \exp\big(-(L^2+z^2)^{1/2}/L_c\big)=\int_L^\infty dR \frac{R}{(R^2-L^2)^{1/2}}\exp(-R/L_c) \cr
=&L K_1(L/L_c) \simeq {\rm constant} \times   (L L_c)^{1/2} \exp(-L/L_c)  ~~~,\cr
\end{align}
with $K_1$ a modified Bessel function and the final expression its large $L$ asymptotic form.  We see that the disk density profile arising from our mechanism is approximately exponential, with disk scale length $L_c \simeq 3.2 {\rm kpc}$. This  matches the observed approximately exponential form of disk galaxy luminosity distributions, with disc scale lengths averaging  $3.8 {\rm kpc}$ with a root mean square dispersion of $2.1{\rm kpc}$ \cite{obs}. It will be interesting to see if accurate observational data can distinguish between a pure exponential, and the Bessel function form $L\,K_1(L/L_c)$  obtained by projecting the original spherical exponential distribution onto a plane.\footnote{Other  mechanisms have been suggested for producing an exponential disk density profile, although without fixing the magnitude of the scale length $L_c$.  A mechanism based on maximizing entropy under angular momentum mixing by radial migration has been suggested by Herpich, Tremaine, and Rix \cite{hpr}, and they give extensive references to earlier proposals. Their mechanism suggests that an initially formed exponential density profile, as in our proposal, would  be stable under subsequent galactic dynamics.}

We have deliberately used the term ``catalyze'' in proposing the action of the central black hole in galactic star formation, because the mass of the central black hole in a typical galaxy is a fraction of a percent of the total galaxy mass.  Hence for our proposal to be viable, either (i) the black hole wind has to nearly compensate for black hole mass accretion, so that most of the galactic mass has passed through the black hole which acts as a catalyst, or (ii) the black hole wind is much smaller than the rate at which the black hole accretes mass, in which case the collisions induced by the wind would have to act as a nucleation for instabilities, eventually involving much more infalling mass, that lead to star formation.  Simulations of galaxy formation incorporating a parameterized black hole wind could test these alternatives.

We suggest several other  calculations to investigate the proposed collision mechanism.  Determining the magnitude and velocity of a possible black hole wind  necessitates solving for the interior structure of the Schwarzschild-like black hole studied in \cite{F}, when matter that makes up its mass is included in the modified gravitational equations.  Studying the effects of black hole rotation requires extending the calculations of \cite{F}, \cite{A2} to ``Kerr-like'' modified black holes.   And seeing whether the proposed mechanism circumvents the problem of excessive angular momentum impeding  star formation \cite{colgate}, involves analyzing  the dynamics of opposing  flows acted on by gravitational forces.
\vfill\eject

To conclude, the close association of black holes with galaxies suggests that galactic core black holes may play a catalytic role in galaxy formation, with the hole seeding star formation, and simultaneously growing in size as the galaxy grows in size.  We have proposed an underlying mechanism for this to occur,  based on the radial geodesic equation for the Schwarzschild metric and the assumption of a ``leaky'' black hole.   The proposed mechanism  can explain the observed exponential luminosity profile of disk galaxies together with the magnitude of the associated scale length, which is related by the formula, ${\it Scale ~Length}\sim (\pi a_0^2 \rho_H)^{-1}$, to the Bohr radius $a_0$ and the density $\rho_H$ of
atomic hydrogen in the proto-galactic region.\footnote{Formation of other types of galaxies, such as elliptical, is believed to involve galactic collisions and mergers, resulting in more complex density profiles.  Elliptical galaxy scale sizes range an order of magnitude in either direction from about 2kpc, which is the same order of magnitude as the mean disk galaxy scale size \cite{tr}.}${}^,$\footnote{Our focus has been on the inner visible baryonic structure of galaxies, and so we have not discussed the dark matter halos that are important in understanding the stellar velocity profiles of the outer regions of galaxies.  Dark matter overdense regions may play a role in formation of the initial black hole.}

\bigskip\bigskip

I wish to thank Kyle Singh for stimulating conversations in the course of writing \cite{K}.  I  thank James Stone for a thought-provoking critique of the initial version of \cite{K}, which led to my investigation of the mechanism reported here, Scott Tremaine for an email giving the reference \cite{hpr} and for useful comments on an initial version of this essay, and Sarah Brett-Smith for asking  whether collisions will occur, which motivated the collision length estimate.

\end{document}